\documentclass[10pt, pra,aps,amsmath,amssymb,superscriptaddress,twocolumn, hypertext,longbibliography]{revtex4-1}
\usepackage{hyperref}
\hypersetup{
    colorlinks,
    linkcolor={black},
    citecolor={blue!80!black},
    urlcolor={blue!80!black}
}
\usepackage{graphicx}
\usepackage{dcolumn}
\usepackage{bm}
\usepackage{xcolor,soul}
\usepackage[utf8]{inputenc}
\usepackage{amsmath}
\usepackage{multirow}
\usepackage{xcolor}
\usepackage[per-mode=symbol]{siunitx}
\DeclareSIUnit\gauss{G}
\graphicspath{{/}}

\newcommand\Ps{\ensuremath{6^2P_{3/2}}}
\newcommand\Ss{\ensuremath{6^2S_{1/2}}}
\newcommand\Ds{\ensuremath{5^2D_{5/2}}}
\newcommand\ND{\ensuremath{N_D}}
\newcommand\NS{\ensuremath{N_S}}
\newcommand\NP{\ensuremath{N_P}}
\newcommand\gD{\ensuremath{\gamma_D}}
\newcommand\gP{\ensuremath{\gamma_P}}
\newcommand\gDS{\ensuremath{\gamma_{D\to S}}}
\newcommand\gDP{\ensuremath{\gamma_{D\to P}}}
\newcommand\tauD{\ensuremath{\tau_D}}
\newcommand\tauP{\ensuremath{\tau_P}}

\newcommand\lifetime{\SI{1353\pm5}{\nano\second}}

\begin{document}


\title{Precise Lifetime Measurement of the Cesium \texorpdfstring{$5^2D_{5/2}$}{5D5/2} State}

\author{S.~Pucher}

\author{P.~Schneeweiss}
\email{philipp.schneeweiss@hu-berlin.de}

\author{A.~Rauschenbeutel}

\affiliation{%
 Vienna Center for Quantum Science and Technology,\\
 TU Wien -- Atominstitut, Stadionallee 2, 1020 Vienna, Austria
}%
\affiliation{%
 Department of Physics, Humboldt-Universit\"at zu Berlin, 10099 Berlin, Germany
}%

\author{A.~Dareau}
\email{alexandre.dareau@institutoptique.fr}
\altaffiliation{\href{https://orcid.org/0000-0001-7581-4701}{orcid.org/0000-0001-7581-4701}}
\affiliation{%
 Vienna Center for Quantum Science and Technology,\\
 TU Wien -- Atominstitut, Stadionallee 2, 1020 Vienna, Austria
}%
\affiliation{%
  Laboratoire Charles Fabry, Institut d'Optique Graduate School, CNRS, Université Paris-Saclay, 91127 Palaiseau cedex, France
}%

\date{\today}

\begin{abstract}
We measure the lifetime of the cesium $5^2D_{5/2}$ state using a time-resolved single-photon-counting method. We excite atoms in a hot vapor cell via an electric quadrupole transition at a wavelength of 685~nm and record the fluorescence of a cascade decay at a wavelength of 852~nm. We extract a lifetime of \lifetime{} for the \Ds{} state, in agreement with a recent theoretical prediction. In particular, the observed lifetime is consistent with the literature values of the polarizabilities of the cesium $6P$ states. Our measurement contributes to resolving a long-standing disagreement between a number of experimental and theoretical results.
\end{abstract}

\maketitle

\section{Introduction}

Alkali atoms, with their simple electronic level structure, provide an ideal test bench for atomic structure theories. Precisely probing alkali atomic properties is, therefore, crucial for the study of fundamental symmetries of the Standard Model. For instance, one of the most precise low-energy tests of parity non-conservation (PNC) in the electroweak interaction was provided by spectroscopic studies of cesium atoms~\cite{Wood97,Porsev09}. Although those measurements have been performed using $S$--$S$ transitions, it has been proposed that $S$--$D$ transitions could be promising candidates to measure PNC effects with an even greater precision~\cite{Roberts14}. In order to compare the results of such experiments with calculations, precise measurements of cesium atomic properties are needed, such as absolute oscillator strengths of atomic transitions. This can be achieved by precisely measuring the lifetimes of cesium $D$ states.

The lifetimes of low-lying $D$ states were measured with a precision of about 1\% in francium~\cite{Grossman00} and rubidium~\cite{Sheng08}. In the case of cesium, the lifetime of the \Ds{} state has been measured with increasing precision for the past forty years~\cite{Marek77,Bouchiat92,Sasso92,Hoeling96,DiBerardino98}, and several models have been developed to make \textit{ab initio} calculations~\cite{Heavens61,Stone62,Warner68,Fabry76,Theodosiou84,Safronova04,Sahoo16,Safronova16} (see Table~\ref{tab:lifetimes_summary} and Fig.~\ref{fig:lifetimes_summary} at the end of the present paper for a summary of those measurements and calculations). The last published measurements by Hoeling \textit{et al.}~\cite{Hoeling96} and DiBerardino \textit{et al.}~\cite{DiBerardino98} report values of \SI{1226(12)}{\nano\second} and \SI{1281(9)}{\nano\second}, respectively. These values disagree clearly beyond their stated error bars. In subsequent theory work, Safronova \textit{et al.}~\cite{Safronova04} showed that there is an inconsistency between the $5^2D_{5/2}$ lifetime values obtained experimentally and the related, independently measured polarizabilities of the $6P$ states of cesium: the expected lifetime of the \Ds{} state, inferred from those experimental polarizability values, amounts to \SI{1359(18)}{\nano\second}~\cite{Safronova04} and was later refined to \SI{1351(52)}{\nano\second}~\cite{Safronova16}. Recently, an independent \textit{ab initio} calculation by Sahoo~\cite{Sahoo16} predicted a lifetime of \SI{1270(28)}{\nano\second}, in agreement with the measurement by DiBerardino \textit{et al.} but in contradiction with the work of Safronova~\textit{et al.}.

In this paper, we report on a precise measurement of the lifetime, $\tauD$, of the \Ds{} state of cesium. We use a standard technique known as time-resolved single-photon-counting. Our measurement yields a lifetime value of \lifetime{}, in agreement with the predictions by Safronova \textit{et al.}~\cite{Safronova04,Safronova16}. The paper is organized as follows: in section II, we present our experimental method and setup; in section III, we discuss the results and the error budget; finally, we present our conclusions in section IV.

\begin{table}[b]
\renewcommand{\arraystretch}{1.1}
\normalsize
    \begin{tabular*}{\columnwidth}{l@{\extracolsep{\fill}}cc}
        \hline
        \hline
        Reference & $\tauD$ (ns) & type \\
        \hline
        Heavens (1961)~\cite{Heavens61} & 1370 & calc. \\
        Stone (1962)~\cite{Stone62} & 1342 & calc. \\
        Warner (1968)~\cite{Warner68} & 1190 & calc. \\
        Fabry (1976)~\cite{Fabry76} & 1434 & calc. \\
        Theodosiou (1984)~\cite{Theodosiou84} & 1283 & calc. \\
        Safronova \textit{et al.} (2004)~\cite{Safronova04} & 1359(18) & calc. \\
        Sahoo (2016)~\cite{Sahoo16} & 1270(28) & calc. \\
        Safronova \textit{et al.} (2016)~\cite{Safronova16} & 1351(52) & calc. \\
        \hline
        Marek \textit{et al.} (1977)~\cite{Marek77} & 890(90) & exp. \\
        Bouchiat \textit{et al.} (1992)~\cite{Bouchiat92} & 1260(80) & exp. \\
        Sasso \textit{et al.} (1992)~\cite{Sasso92} & 1250(115) & exp. \\
        Hoeling \textit{et al.} (1996)~\cite{Hoeling96} & 1226(12) & exp. \\
        DiBerardino \textit{et al.} (1998)~\cite{DiBerardino98} & 1281(9) & exp. \\
        \hline
        this work & \lifetime{} & exp. \\
        \hline
        \hline
    \end{tabular*}
    \caption{Review of calculations (calc.) and experimental results (exp.) of the \Ds{} state lifetime \tauD{} found in the current literature (see also Fig.~\ref{fig:lifetimes_summary}).}
    \label{tab:lifetimes_summary}
\end{table}

\section{Experimental method and setup}

\subsection{Atomic Structure}

\begin{figure}[h]
    \centering
    \includegraphics[width=\columnwidth]{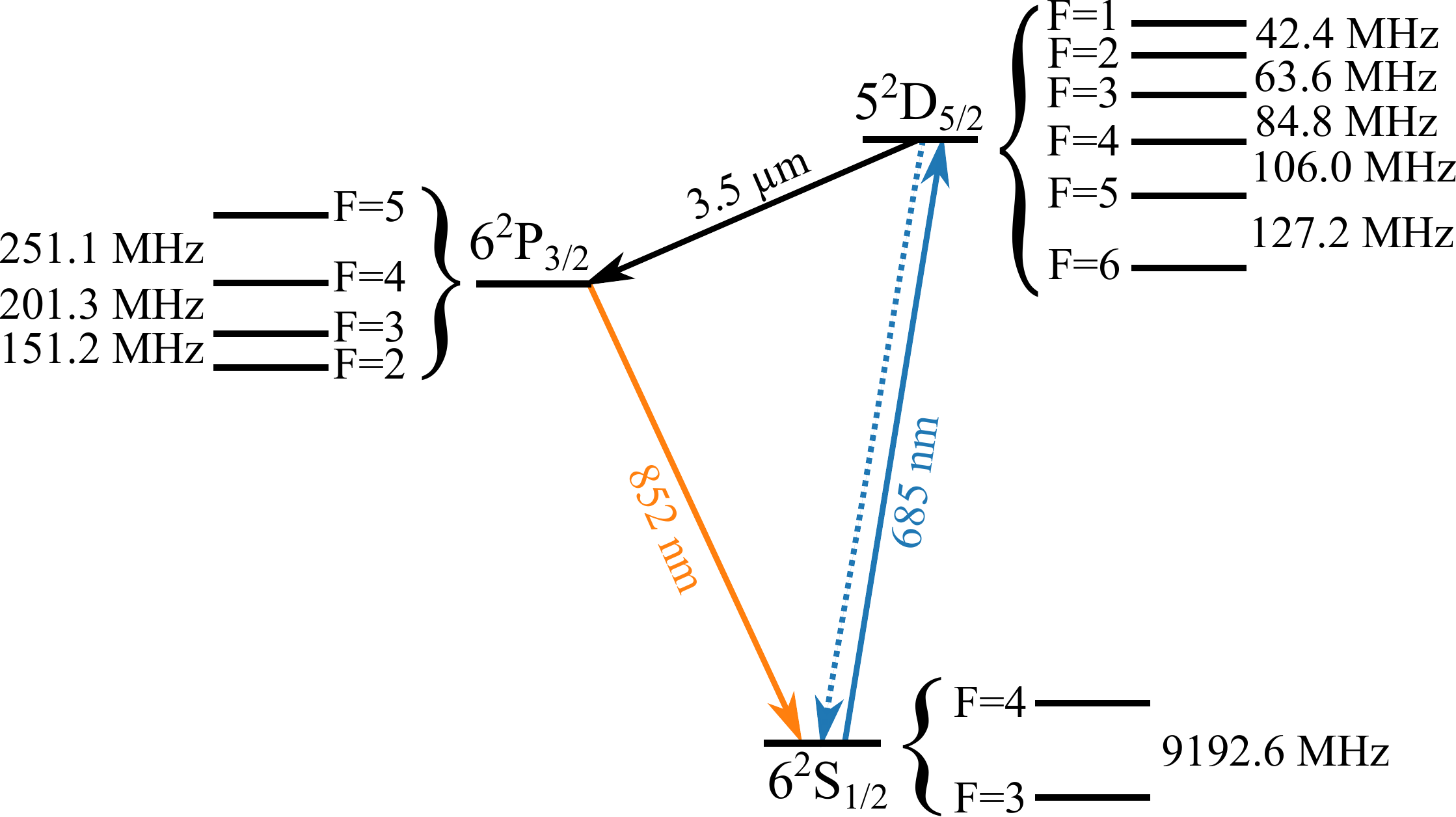}
    \caption{Cesium energy levels relevant to our experiment. We excite atoms via an electric quadrupole transition at a wavelength of \SI{685}{\nano\meter}, and we detect the fluorescence emitted at a wavelength of \SI{852}{\nano\meter}.}
    \label{fig:atomic_structure}
\end{figure}

The energy levels of cesium relevant for the present measurement are shown in Fig.~\ref{fig:atomic_structure}. We excite atoms from the \Ss{} ground state to the \Ds{} state via an electric quadrupole transition at a wavelength of \SI{685}{\nano\meter}. From this excited state, most of the atoms decay to the \Ps{} state via an electric dipole transition at a wavelength of \SI{3.5}{\micro\meter}. The atoms will dwell, on average, $\tauP=\SI{30.462(46)}{\nano\second}$~\cite{Patterson15} in this intermediate state, and then decay back to the ground state. We detect the fluorescence photons from the last transition at a wavelength of \SI{852}{\nano\meter}.

\subsection{Laser Setup}

\begin{figure}[b]
    \centering
    \includegraphics[width=0.9\columnwidth]{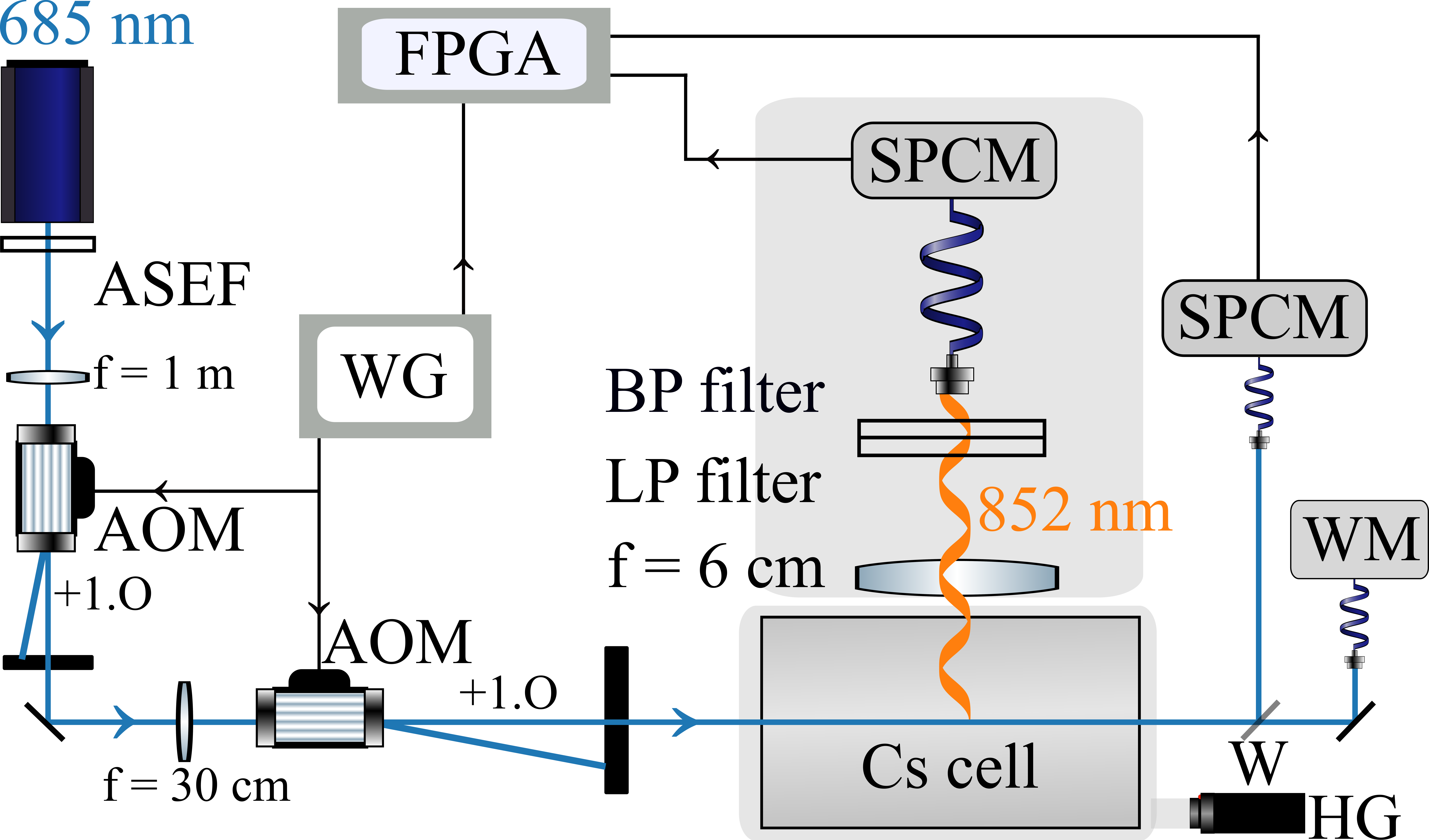}
    \caption{Diagram of our experimental setup. We send a laser beam at a wavelength of \SI{685}{\nano\meter} through two acousto-optic modulators (AOM), and then to a cell containing cesium vapor. A single-photon counting module (SPCM) detects fluorescence photons. ASEF: amplified spontaneous emission filter, BP filter: bandpass filter, LP filter: longpass filter, WG: waveform generator, FPGA: field-programmable gate array, WM: wavelength meter, HG: heating gun.}

    \label{fig:experimental_setup}
\end{figure}

Our experimental setup is illustrated in Fig.~\ref{fig:experimental_setup}. Light at a wavelength of \SI{685}{\nano\meter} is obtained using a tapered amplifier laser, which we will later refer to as the excitation laser. Directly after the output of the laser, a shortpass filter~\footnote{Thorlabs, FESH0800} with a cut-off wavelength of \SI{800}{\nano\meter} suppresses the amplified spontaneous emission of the laser around the fluorescence wavelength at \SI{852}{\nano\meter}.

The excitation laser beam is then sent through two acousto-optic modulators (AOM) using the first diffraction order in both cases. By switching the RF power supplied to the AOMs, the excitation beam can be turned on and off. The beam is sent to a commercial spectroscopy cell~\footnote{Thorlabs, GC25075-CS} containing a hot cesium vapor. The excitation laser beam has a power of about \SI{21}{\milli\watt} and a beam diameter of about \SI{1.2}{\milli\meter} in front of the vapor cell. This corresponds to an intensity of \SI{2.3}{\watt/\centi\meter^2}, which is on the order of the saturation intensity of the quadrupole transition~\cite{Chan16}.

The laser light that is transmitted through the cell is detected using a single photon counting module (SPCM,~\footnote{Excelitas Technologies, SPCM-AQRH-14-FC}).
This reference SPCM allows us to monitor the switch-off behavior of the laser (see Fig.~\ref{fig:fluorescence_decay_fit}). We observe a delay of about \SI{500}{\nano\second} between the electronic signal commanding the excitation laser beam turn-off and the actual intensity decay. Then, the laser beam is suppressed to about \SI{0.1}{\percent} of its initial power in a few tens of nanoseconds.

In order to tune the laser into resonance with the $\Ss\to\Ds$ transition, we measure the fluorescence signal from the vapor cell as a function of the laser frequency. The latter is measured by sending part of the excitation beam to a wavelength meter~\footnote{HighFinesse, WS-6}. We obtain a Doppler-broadened signal with a full width at half maximum of \SI{477 \pm 4}{\mega\hertz}, which allows us to resolve the hyperfine structure of the \Ss{} ground state. We perform a frequency scan at the beginning of each experimental run and determine the laser setting in order to excite the $\Ss(F=4)\to \Ds$ transition. The laser frequency is then kept constant during the entire experimental run (consisting of several cycles, as explained in the next section) using the wavelength meter. We checked that drifts of the wavelength meter are irrelevant under our lab conditions and, thus, do not give rise to an error in the frequency stabilization of the excitation laser.

All the measurements shown in the present paper are performed with the excitation laser tuned to the $\Ss(F=4)\to \Ds$ transition. Due to Doppler broadening, we cannot resolve the hyperfine structure of the \Ds{} state. Thus, we excite atoms to all the hyperfine $F$-states. Since all those states are expected to have the same lifetime, this has no effect on the outcome of our measurement.

\subsection{Fluorescence Setup}

We use a commercial cesium spectroscopy cell made of borosilicate glass. The cell has a length of \SI{7.18}{\centi\meter} and a diameter of \SI{2.54}{\centi\meter}, and is enclosed inside a metal box. We set the temperature of the cell by letting heated airflow through the box.

The fluorescence light emitted by the atoms into a direction perpendicular to the excitation laser beam path is collected using a lens with a focal length of \SI{6}{\centi\meter} and a diameter of \SI{5.08}{\centi\meter}. A longpass filter~\footnote{Semrock, BLP01-808R-25} with a cut-off wavelength of \SI{808}{\nano\meter} is used to suppress stray light at a wavelength of \SI{685}{\nano\meter} that stems from the scattering off the various optical interfaces. A bandpass filter~\footnote{Semrock, LL01-852-12.5}, centered at \SI{852}{\nano\meter}, further reduces background photon counts. The filtered fluorescence light is then sent onto an SPCM via a multimode optical fiber. The arrival times of the detected fluorescence photons and those of the reference SPCM signal are recorded using a field-programmable gate array (FPGA, \footnote{Opal Kelly, XEM3005-1200M32P}).

\subsection{Experimental Sequence}

In an excitation cycle performed at room temperature, the excitation laser beam is switched on for \SI{25}{\micro\second}, so that a steady state of the fluorescence signal is reached. The excitation is then switched off and stays off for \SI{25}{\micro\second}, in order to measure the decay of the fluorescence. This cycle is then repeated until good counting statistics is reached. At higher temperatures, radiation trapping of photons emitted at \SI{852}{\nano\meter} slows down the fluorescence dynamics. Thus, we use longer excitation cycles.

\begin{figure}[t]
    \centering
    \includegraphics[width=\columnwidth]{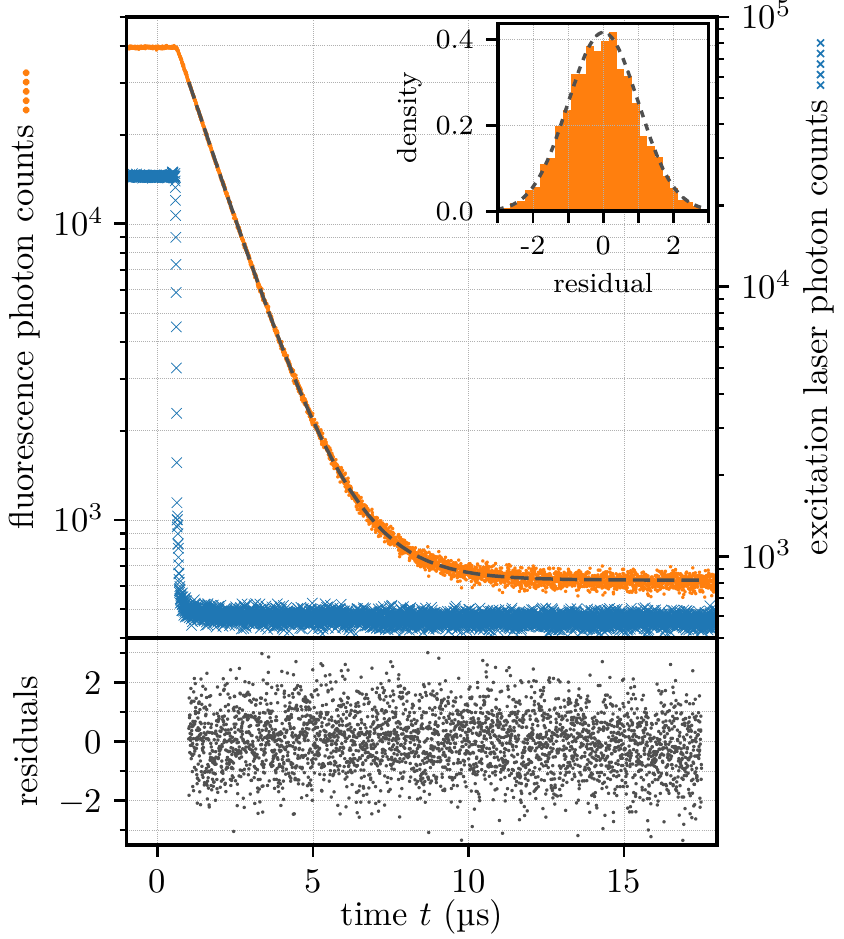}
    \caption{Top panel: typical decay curves for the atomic fluorescence (orange dots) and the excitation laser (blue crosses) intensities. We send the trigger commanding the laser to switch off at $t=\SI{0}{\micro\second}$. We detect the signals with two separate SPCMs, and store the photon arrival times. From this data, we generate the photon count histograms using a bin size of \SI{5}{\nano\second}. We show the result of a fit using a single exponential decay with an offset as a dashed gray line. The distribution of the normalized fit residuals (see main text for details) is shown as an inset (orange bars). The dashed gray line in the inset corresponds to a Gaussian distribution centered on zero with a variance of one. Bottom panel: normalized fit residuals for each time bin.}
    \label{fig:fluorescence_decay_fit}
\end{figure}

We store the arrival times of photons detected by the fluorescence and the reference SPCMs using the FPGA and also record the electronic signal triggering the laser switch-off. We then compute the delay between the arrival time of each detected photon and the beginning of the respective experimental cycle. Their histogram is shown in Fig.~\ref{fig:fluorescence_decay_fit} (orange dots); note the logarithmic scale of the ordinate axis. The fluorescence signal starts at \SI{4e4}{counts/bin}, and shows an exponential decay until background counts (about \SI{600}{counts/bin}) begin to dominate. Moreover, a sharp switch-off behavior for the excitation laser light is observed (blue crosses). A typical experimental run consists of about $10^9$ excitation cycles and takes about 14 hours. We detect a total of about $10^8$ fluorescence photons during one measurement series, which corresponds to a rate of about $0.1$ detected photons per excitation cycle.

\section{Results}

\subsection{Extracting the Lifetime via a Fit}

\subsubsection{Theoretical Model}
\label{sec:theo_model}

We now describe the model we use to fit the fluorescence decay shown in Fig.~\ref{fig:fluorescence_decay_fit}. We consider the three-level system depicted in Fig.~\ref{fig:atomic_structure}. In a typical fluorescence measurement, atoms are illuminated with laser light resonant with the electric quadrupole $\Ss (F=4) \to \Ds$ transition. After \SI{25}{\micro\second} of illumination, the laser is switched off. Populations in the different states can be calculated by solving the following rate equations:
\begin{align}
\dot{N}_D & =  -\left(\gDP + \gDS \right) \ND + P \NS - P \ND, \label{eq:rates_D}\\
\dot{N}_P & =  \gDP \ND - \gP \NP,\label{eq:rates_P}\\
\dot{N}_S & = \gDS \ND + \gP \NP - P\NS + P\ND, \label{eq:rates_S}
\end{align}

\noindent where \ND{}, \NP{}, and \NS{} are the respective populations in the \Ds{}, \Ps{}, and \Ss{} states, and $P$ is a pumping rate depending on the laser settings. The decay rate on the electric quadrupole transition is $\gDS \approx 2\pi\times\SI{3.5}{\hertz}$, the decay rate for the $\Ds\to\Ps$ transition is $\gDP \approx 2\pi\times\SI{124}{\kilo\hertz}$, and the decay rate on the $\Ps\to\Ss$ transition is $\gP \approx 2\pi\times\SI{5.2}{\mega\hertz}$. Since $\gDS \ll \gDP$, we can neglect direct fluorescence on the electric quadrupole transition. The total decay rate from the \Ds{} state, \gD{}, is then dominated by the decay to \Ps{}, so that $\gD = \gDP + \gDS \approx \gDP$.

In our experiment, we make sure that the excitation laser beam is switched on for a sufficiently long time to reach a steady state of the fluorescence intensity $I_0$. Starting from this steady state, we obtain the time-resolved fluorescence intensity, $I(t)$, when the laser is switched off ($P=0$) by solving equations~(\ref{eq:rates_D}-\ref{eq:rates_S}). We find:
\begin{equation}
I(t) = I_0 \left\{  \frac{\gP}{\gP - \gD} e^{-\gD t} - \frac{\gD}{\gP - \gD} e^{-\gP t}\right\}.
\label{eq:decay_formula}
\end{equation}

Thus, the fluorescence signal is the sum of two exponential terms with respective decay rates \gD{} and \gP{}. In our case, since $\gP \gg \gD$, the contribution of the second term to the fluorescence signal is already small at $t=0$ (about \SI{2}{\percent} of the total intensity). Moreover, the second term decays much faster than the first one: for instance, \SI{500}{\nano\second} after the laser switch-off, this term only contributes to about $2\times 10^{-9}$ of the total intensity. As explained later, we start fitting the fluorescence signal at least \SI{500}{\nano\second} after the laser switch-off. We can, therefore, safely neglect the second term of eq.~\eqref{eq:decay_formula}, and fit our data only taking into account a single exponential decay.

\subsubsection{Fit Method}
\label{sec:fit_method}

We fit the raw data using the LMFIT python package~\footnote{\url{https://lmfit.github.io/lmfit-py/}}, which relies on a non-linear least-squares minimization method. We use the following fit formula:
\begin{equation}
    I^\mathrm{(fit)}(t) = I_0 \exp\left(-\frac{t}{\tau_D}\right) + c,
    \label{eq:fitmodel}
\end{equation}

\noindent where the free fit parameters are the initial intensity, $I_0$, the decay time constant, $\tau_D$, and a constant offset, $c$, which takes background photons and detector dark counts into account. In order to check the goodness of the fit, we consider the normalized fit residuals (see lower panel of Fig.~\ref{fig:fluorescence_decay_fit}), defined as:
\begin{equation}
r_i = \frac{y_i - y_i^{(\mathrm{fit})}}{\sigma_i} = \frac{y_i - y_i^{(\mathrm{fit})}}{\sqrt{y_i}},
\end{equation}

\noindent where $y_i$ ($y_i^{(\mathrm{fit})}$) are the measured (fitted) photon counts in the $i^\mathrm{th}$ time bin, and $\sigma_i$ is the standard deviation of the variable $y_i$ (we assume a Poisson distribution, hence $\sigma_i = \sqrt{y_i}$). Since the number of time bins, $N$, is large compared to the number of free parameters in our fit model, the reduced chi-squared $\tilde{\chi}^2$ equals the mean of the squared residuals: $\tilde{\chi}^2 = (\sum_{i=1}^{N} r_i^2) / N$. A reduced chi-squared close to unity is an indicator of a good fit. For the fit shown in Fig.~\ref{fig:fluorescence_decay_fit}, we obtain $\tau_D = \SI{1353.2\pm0.5}{\nano\second}$, while $\tilde{\chi}^2 = 1.0009$.
Here, the error given for the fit result corresponds to the 68\% confidence interval.
We also checked that the distribution of the normalized residuals follows a Gaussian (see the upper panel of Fig.~\ref{fig:fluorescence_decay_fit}, inset). This is the case, providing an additional indication that our model is adequate for fitting our experimental data.

\subsubsection{Fit Results and Fluctuations}

\begin{figure}[ht]
    \centering
    \includegraphics[width=\columnwidth]{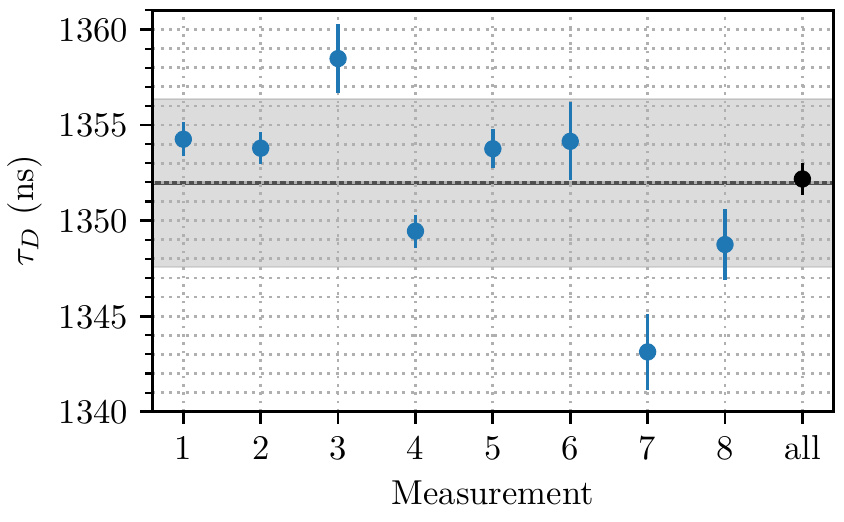}
    \caption{ Fitted lifetime for the \Ds{} state for six independent measurement runs (see main text for more details). The result of a fit on the combined data is also shown (black circle). The error bars take into account the fit 68\% confidence interval and the truncation error, \textit{i.e.}, the standard deviation of the fit result when varying the fitting range (cf.~section~\ref{sec:trunc_error}). The average of the eight lifetime measurements is marked by a solid black line, and the standard deviation is indicated by a shaded area. Note that the solid line is obtained by averaging the individual fits on measurements 1 to 8, while the black circle corresponds to a single fit on a signal generated by combining the photon arrival times from all those measurements. The lifetimes obtained by those two methods are thus slightly different.}
    \label{fig:fit_statistics}
\end{figure}

A first estimate of the error of the lifetime inferred from the fit of the fluorescence decay is given by the 68\% confidence interval provided by the fit routine. This confidence interval is about \SI{1}{\nano\second} for the lifetime of the \Ds{} state. It is consistent with an independent estimation made using the bootstrap method~\cite{Efron94}. However, this confidence interval seems to underestimate fluctuations in our measurements. This was confirmed by performing several independent measurement runs and comparing their results. The outcome of our entire measurement campaign is shown in Fig.~\ref{fig:fit_statistics}. Each measurement run had approximately the same duration and the same photon count rate. They were taken over a period of about eight weeks during six experimental runs. Data points 1 \& 2 and 5 \& 6 issue from longer measurement runs which were divided into two subsets of equal sizes. Measurements were taken at temperatures ranging from \SI{22.5}{\celsius} to \SI{24}{\celsius}, with a stability of about \SI{0.4}{\celsius} within one measurement. Averaging the eight measured values, we obtain a mean lifetime of \SI{1352.0}{\nano\second}, with a standard deviation of \SI{4.4}{\nano\second}. A fit on the data obtained by combining all measurements yields a lifetime of \SI{1353.2\pm0.3}{\nano\second}.

\subsection{Systematic Errors}
\label{sec:errors}
We now discuss systematic errors that can impact our lifetime measurement.

\subsubsection{Radiation Trapping}

Since the population of atoms in the \Ps{} state is small, we expect no radiation trapping for photons emitted at a wavelength of \SI{3.5}{\micro\meter}, cf. Fig.~\ref{fig:atomic_structure}. However, radiation trapping can occur for the photons subsequently emitted at a wavelength of \SI{852}{\nano\meter}. This might increase the apparent lifetime of the \Ps{} state. The fluorescence signal can then be expressed as the sum of contributions from a series of eigenmodes (the so-called Holstein modes) which decay exponentially with a time constant $\tau_P^{(i)} = g_i \tau_P$, where $i$ is a mode index and $g_i$ the Holstein radiation trapping factor for a Doppler broadened ensemble of atoms~\cite{Molisch93}. This factor depends both on the geometry of the cell and on the absorption coefficient $\alpha(\delta)=n \sigma(\delta)$ where $n$ is the atomic number density and $\sigma(\delta)$ is the laser detuning-dependent absorption cross section averaged over all polarizations. For a cylindrical cell of radius $r$, the first Holstein factors were conveniently expressed as a function of the attenuation parameter $\alpha r$ in Ref.~\cite{Molisch93}. For the measurements carried out at a temperature of \SI{23}{\celsius}, we expect $n\approx\SI{3.4e10}{cm^{-3}}$, corresponding to $\alpha r \approx 0.66$. For this setting, the largest Holstein factor is $g_1=1.6$, thus corresponding to a 60\% increase of the \Ps{} state lifetime.
For the considered fitting ranges (cf. section~\ref{sec:theo_model}), the contribution of the \Ps{} state is therefore negligible, even in the presence of radiation trapping.

\subsubsection{Effect of Atomic Collisions}
\label{sec:collisions}
\begin{figure}[tb]
    \centering
    \includegraphics[width=\columnwidth]{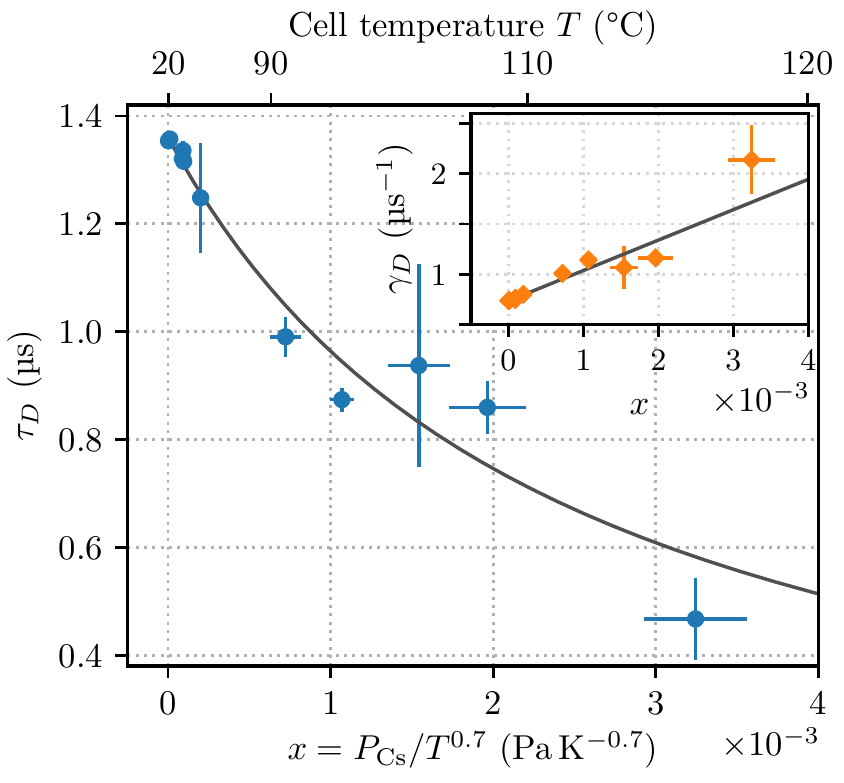}
    \caption{Inset: fitted decay rate, $\gamma_D = \gamma_\mathrm{D,0} + \gamma_{D, \mathrm{coll}}$, of the \Ds{} state as a function of $x=P_\mathrm{Cs}/T^{0.7}$ (see main text). As expected, the collisional decay rate scales linearly with $x$. A linear fit (solid line) yields a slope of \SI{300\pm30}{\micro\second^{-1}.Pa^{-1}.K^{0.7}}. Main panel: Lifetime, $\tau_D = 1/\gamma_D$, of the \Ds{} state as a function of $x$. The solid line corresponds to the linear fit on the decay rate shown in the inset. The cell temperature, $T$, corresponding to the considered value of $x$ is given at the top of the figure. For the two panels, error bars take into account both the fit 68\% confidence interval and the truncation error (cf.~section~\ref{sec:trunc_error}).}
    \label{fig:systematics_temperature}
\end{figure}

Inelastic collisions between cesium atoms might reduce the lifetime of excited state and, therefore, influence our measurements. The collision rate depends on $n$ as well as on the vapor temperature. According to the Lindholm-Foley model, the total decay rate of the \Ds{} state $\gamma_D$ can be written as~\cite{Fletcher95,Dareau15}:
\begin{equation}
 \gamma_D = \gamma_\mathrm{D,0} + \gamma_{D, \mathrm{coll}}~,
 \end{equation}

 \noindent where $\gamma_\mathrm{D,0}$ is the natural decay rate and $\gamma_{D, \mathrm{coll}}$ is the collision-induced decay rate.  The latter can be written as:

\begin{equation}
 \gamma_{D, \mathrm{coll}} = A \times \frac{P_\mathrm{Cs}}{T^{0.7}}~,
\end{equation}

\noindent where $T$ is the temperature of the cesium vapor, $P_\mathrm{Cs}$ is the partial cesium pressure in the cell, and $A$ is a constant which needs to be determined. Fig.~\ref{fig:systematics_temperature} shows $\gamma_D$ measured for different values of the parameter $x=P_\mathrm{Cs}/T^{0.7}$. The latter is varied by changing $T$, which also results in a variation of $P_\mathrm{Cs}$. A linear fit yields $A=\SI{300\pm30}{\micro\second^{-1}.Pa^{-1}.K^{0.7}}$. The measurements shown in Fig.~\ref{fig:fit_statistics} were carried out at a mean temperature of \SI{23}{\celsius}, with a dispersion of about \SI{1}{\celsius}. At this temperature, we expect a partial pressure of \SI{1.4e-6}{\milli\bar}, which corresponds to $x=\SI{2.6e-6}{Pa.K^{-0.7}}$. This yields an estimated collision-induced decay rate $\gamma_{D,\mathrm{coll}} = \SI{7.8\pm0.8e-4}{\micro\second^{-1}}$. Therefore, the total lifetime of \SI{1352}{\nano\second} features a systematic error of \SI{1.4\pm0.1}{\nano\second} due to collisions. Finally, the temperature dispersion of about \SI{1}{\celsius} in our measurements yields an additional uncertainty of \SI{0.1}{\nano\second}, which we add to our error budget.

\subsubsection{FPGA Accuracy}

In order to check the accuracy of the FPGA clock, we compare it to a reference signal from a \SI{10}{\mega\hertz} rubidium-based atomic clock~\footnote{Stanford Research Systems, FS725}. The clock was connected to the FPGA, which was set to record one time-tag per clock period. We then computed the distribution of time delays between two consecutive time-tags. We find a mean deviation of \SI{5.9(1)}{\pico\second} from the expected time delay, which corresponds to a relative error of \SI{6e-5}{}. For our measured lifetime values, this yields an error of about \SI{90}{\pico\second}, which is negligible in our error budget. The FWHM of the measured time delay distribution is about \SI{78.0(1)}{\pico\second}, which is also negligible in our final error budget.

\subsubsection{Truncation Error}
\label{sec:trunc_error}
The truncation error stands for the variation of the inferred lifetime that arises when varying the fitting interval. For our fit, we only take into account data that was recorded at least \SI{500}{\nano\second} after the actual laser switch-off. This is done in order to neglect the contribution of the \Ps{} state lifetime in the fluorescence signal. In order to find the optimum start and stop times for the fit, we ran the fit while scanning the fitting range in a two-dimensional way. For each iteration, we checked the reduced chi-squared and the 68\% confidence interval for the fit results. We find that the fit works well for start and stop times ranging from \SI{0.9}{\micro\second} to \SI{1.3}{\micro\second} and \SI{13}{\micro\second} to \SI{17}{\micro\second} after the laser switch-off, respectively. We then estimate the truncation error by computing the standard deviation of the fit result when varying the start and stop time within this range. For the measurements shown in Fig.~\ref{fig:fit_statistics}, we obtain a mean truncation error of \SI{0.9}{\nano\second}, with a mean reduced chi-squared of \SI{1.01\pm0.02}{}.

\subsubsection{SPCM Dead Time}

Our SPCM is specified to have a dead time of \SI{22}{\nano\second}. Therefore, if a photon arrives on the SPCM less than \SI{22}{\nano\second} after the detection of another photon, it will not be detected. This alters the distribution of the arrival times when the photon flux is large. In our measurements, we detect, on average, $0.1$ photons per \SI{50}{\micro\second} cycle, corresponding to an average delay of \SI{500}{\micro\second} between two consecutive photons. This is much larger than the SPCM dead time, and we can neglect this effect.

\subsubsection{Quantum Beats}
The presence of multiple decay paths from the \Ds{} excited state to the \Ss{} ground state might lead to the appearance of quantum beats~\cite{Silverman78} in the fluorescence signal. The beat frequency is set by the difference of optical transition frequencies of the states involved in the fluorescence process. Beats originating from different excited-state hyperfine levels would have a frequency of tens to hundreds of MHz, i.e., corresponding to an oscillation period much shorter than the \Ds{} state lifetime. However, quantum beats originating from transitions from/to different Zeeman states can have a period comparable to the investigated d-state lifetime for typical stray magnetic fields. This can lead to a systematic error in the lifetime measurement.

When fitting our data using Eq.~\ref{eq:fitmodel}, such a beat would be visible in the fit residuals (see Fig.~\ref{fig:fluorescence_decay_fit}). For all measurements that are presented in this paper, we checked that no such beat was present by analyzing the Fourier transform of the fit residuals. Furthermore, the frequency of quantum beats between different Zeeman states is expected to depend on the magnetic field present at the atoms. In order to check for this effect, we performed measurements at different magnetic fields up to 25~G, using a coil that encloses the cesium cell. No quantum beats were observed. In summary, we conclude that the effect of quantum beats can be neglected for lifetime estimation.

\subsubsection{Other Systematics}

Other systematic effects, such as wall collisions~\cite{DiBerardino98,Hoeling96}, blackbody radiation~\cite{Hoeling96}, afterpulses~\cite{Hoeling96} or pulse pileup correction~\cite{Sheng08} should be negligible in our experimental configuration.

\begin{table}[htb]
\renewcommand{\arraystretch}{1.1}
    \begin{tabular*}{\columnwidth}{l@{\extracolsep{\fill}}rr}
        \hline
        \hline
        Source & Correction (ns)  & Error (ns)\\
        \hline
        Fit confidence interval & & 1.0 \\
        Statistical (cf. Fig.~\ref{fig:fit_statistics}) & & 4.4 \\
        Collisional broadening & +1.4 & 0.1 + 0.1 \\
        FPGA accuracy &  & $< 0.1$ \\
        SPCM dead time &  & $< 0.1$ \\
        Quantum beats &  & $< 0.1$ \\
        Truncation error &  & 0.9 \\
        \hline
        Total & +1.4 & 4.6\\
        \hline
        \hline
    \end{tabular*}
    \caption{Summary of the relevant corrections and errors considered for the determination of the \Ds{} state lifetime. For the collisional broadening, there are two contributions arising from the temperature uncertainty and from the uncertainty in the measurement of the collision-induced decay rate, see section~\ref{sec:collisions}.}
    \label{tab:error_budget}
\end{table}

\section{Conclusion}

\begin{figure}[htb]
    \centering
    \includegraphics[width=\columnwidth]{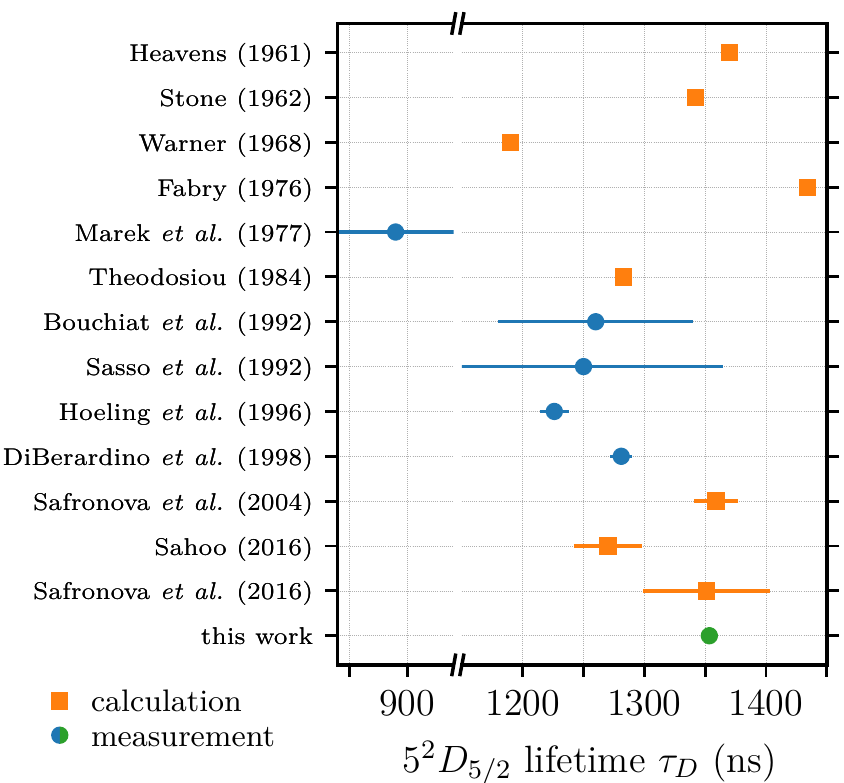}
    \caption{Review of calculations (yellow squares) and measurements (blue circles) of the \Ds{} state lifetime \tauD{} found in the current literature. The value measured in the present work is marked by a green circle. Lifetime values and corresponding references are listed in Table~\ref{tab:lifetimes_summary}.}
    \label{fig:lifetimes_summary}
\end{figure}

The final error budget for our measurement is summarized in Table~\ref{tab:error_budget}. Taking into account all errors and corrections, we obtain a final value of \lifetime{} for the lifetime of the \Ds{} state. The comparison of our result with previous measurements and calculations is provided in Fig.~\ref{fig:lifetimes_summary}. Our value disagrees with the measurements by Hoeling~\textit{et al.}~\cite{Hoeling96} and DiBerardino~\textit{et al.}~\cite{DiBerardino98}, while it agrees with the calculations by Safronova~\textit{et al.}~\cite{Safronova04,Safronova16}. Thus, it should also be consistent with known values of the cesium $6P$ state polarizabilities. Note that our measurement is not compatible with the calculation of Sahoo~\cite{Sahoo16}, which is consistent with the values measured by DiBerardino~\textit{et al.}.\\
We hope that our results serve future theoretical predictions of cesium's atomic properties, and will help to resolve the current discrepancy between independent calculations of the \Ds{} state lifetime. A modification of our experimental setup would allow us to extend this study to the lifetime of the cesium $5^2D_{3/2}$ state, which can be excited using a laser with a wavelength of \SI{689}{nm}. For this state, the measured~\cite{DiBerardino98} and calculated~\cite{Safronova04,Sahoo16} lifetime values in the literature also disagree. Overcoming these inconsistencies and further improving the knowledge of the electronic structure of cesium will aid the test of parity non-conservation and be beneficial for fundamental studies of atomic physics in general.

\section*{Acknowledgments}

We thank L.~Orozco for stimulating discussions and helpful comments. Financial support from the European Union’s Horizon 2020 research and innovation
program under grant agreement No.~800942 (ErBeStA), the European Research Council (CoG NanoQuaNt), as well as from the Austrian Academy of Sciences (ÖAW, ESQ DiscoveryGrant QuantSurf) is gratefully acknowledged.

\section*{Data availability}

In order to improve traceability of the presented measurements and analysis, the authors commit to publishing the raw experimental data used in this paper in an open-access repository. In addition, the source codes used for the analysis of the data will be shared. A link to the data will be provided in the final, peer-reviewed version of this manuscript.


\bibliography{bibliography}

\end{document}